\documentstyle[12pt,fleqn]{article}
\newcommand{\preprint}[1]{\hfill{\sl preprint - #1}\par\bigskip\par\rm}
\def\titolo{\par\bigskip\begin{center}\bf\LARGE}
\def\endtitolo{\end{center}\par\bigskip\par\rm\normalsize}
\def\instit{\begin{center}\large}
\def\endinstit{\end{center}\rm\normalsize}
\def\references{\begin{thebibliography}{10}}
\def\endreferences{\end{thebibliography}\end{document}}
\newcommand{\dip}{\smallskip Dipartimento di Fisica,
                                Universit\`a di Trento}
\newcommand{\infn}{\smallskip Istituto Nazionale di Fisica Nucleare,\\
                                 Gruppo Collegato di Trento,\\ 38050 Povo (TN)
Italia}  
\newcommand{\dinfn}{\dip\\ and \infn}
\newcommand{\btit}{\begin{titolo}}
\newcommand{\etit}{\end{titolo}}

\newcommand{\Idinfn}{\begin{instit}\dinfn\end{instit}}

\renewcommand{\author}[1]{\begin{center}\Large #1\end{center}}
\renewcommand{\date}[1]{\par\bigskip\par\sl\hfill #1\par\medskip\par\rm}
\newcommand{\email}[1]{e-mail: \sl #1@alpha.science.unitn.it \rm}
\newcommand{\femail}[1]{\footnote{\email{#1}}}
\newcommand{\pacs}[1]{\smallskip\noindent{\sl PACS number(s):
                       \hspace{0.3cm}#1}\par\bigskip\rm}
\newcommand{\babs}{\hrule\par\begin{description}\item{Abstract: }\it}
\newcommand{\eabs}{\par\end{description}\hrule\par\medskip\rm}
\newcommand{\ack}[1]{\par\section*{Acknowledgements} #1}
\renewcommand{\vec}[1]{{\bf #1}}       
\newcommand{\nn}{\nonumber}            
\newcommand{\beq}{\begin{eqnarray}}    
\newcommand{\eeq}{\end{eqnarray}}      
\newcommand{\beqn}{\begin{eqnarray}}   
\newcommand{\eeqn}{\end{eqnarray}}     
\newcommand{\at}{\left(}               
\newcommand{\ct}{\right)}              
\newcommand{\R}{\mbox{$I\!\!R$}}                 
\newcommand{\ii}{\infty}                         
\newcommand{\al}{\alpha}
\newcommand{\be}{\beta}

\newcommand{\de}{\delta}
\newcommand{\ep}{\varepsilon}

\newcommand{\si}{\sigma}
\newcommand{\om}{\omega}

\newcommand{\Om}{\Omega}

\begin{document}


\preprint{UTF 350 }
\btit
  Wightman Functions' Behaviour
  on the Event Horizon \\ of an Extremal Reissner-Nordstr\"{o}m Black Hole.
\etit

\author{Valter Moretti \femail{moretti}}

\Idinfn

\date{May - 1995}

\babs \\
 A weaker  Haag, Narnhofer and Stein  prescription as well as a
 weaker   Hessling Quantum Equivalence Principle for the behaviour 
 of thermal Wightman functions on an event horizon are analysed 
 in the case of an  extremal Reissner-Nordstr\"{o}m black hole 
 in the limit of a large mass. In order to avoid the degeneracy  
 of the metric in the stationary coordinates on the horizon, 
 a method is introduced which employs the 
 invariant length  of geodesics which pass the horizon.\\ 
 First the method is checked for a massless scalar field on the 
 event horizon of the  Rindler wedge, extending the original procedure
 of Haag, Narnhofer and Stein onto the {\em whole horizon} and recovering
 the same results found by Hessling.\\
  Afterwards the  HNS  prescription and  Hessling's prescription 
 for a massless scalar field are
 analysed on the {\em whole horizon} of an extremal Reissner-Nordstr\"{o}m 
 black hole in the limit of a large mass.
It is proved that the weak form of the  HNS prescription is satisfyed for all
 the finite
 values of the temperature of the KMS states, i.e., this principle does not
determine any Hawking temperature. 
 It is found that the Reissner-Nordstr\"{o}m vacuum, i.e., $T=0$ does
 satisfy the weak HNS prescription and it is the only state which 
 satisfies weak Hessling's prescription, too.
  Finally it is suggested that all the previously obtained results should be
  valid dropping the requirements of a massless field and of a large mass 
 black hole, too.
\eabs 



 \section*{Introduction}

 Recently Hawking, Horowitz and Ross \cite{H-H} have discussed the 
 thermodynamics of
 an extremal Reissner-Nordstr\"{o}m black hole.
 It seems to follow from this discussion that thermal
 states in thermal equilibrium
 outside of it  may have every value of temperature because the extremal R-N
 black hole may have every value of temperature, too.\\
 In literature there exist different methods of seeking the possible 
 temperatures
 of thermal states of free scalar fields 
 in a  static globally hyperbolic space-time region with horizons.
 The most common and popular method 
 consists of two mathematical steps (see for example \cite{prolungamento}).\\
 At the beginning one  has to extend the time coordinate to imaginary values
 and  eliminate all the metric singularities connected to the horizon by an 
 opportune choice
  of the imaginary time periodicity $\beta_{M}$.
 The second step is to impose the  KMS condition for thermal 
 states
 \cite{kms,libro haag}, i.e.,
 to impose the  periodicity condition on the 
 imaginary time dependence  of the thermal
 Wightman functions and interpret the common 
  period 
 $\be_{T}$ as $1/T$,
 where T is the temperature of the state. Note that, because of
 the time periodicity  of the manifold,  it is not possible to fix 
 the value of $\be_{T}$ arbitrarily, but the permitted values must be
 of the form $\be_{n}=\be_{M}/n$, where $n= 1,2,3...$.
 The value obtained for $n=1$, i.e., the periodicity of the imaginary time 
 manifold  $\be_{1}=\be_{M}$  determines the Hawking temperature:
 $T_{H}=1/\be_{1}=1/\be_{M}$.\\
 However it is important to stress that the integers  $n=2,3,4...$
 produce also  correct 
 periodic Wightman functions on the same
 Euclidean manifold (see \cite{gorini} on similar topics)
 and {\em a priori} there is  no reason to reject these
 additional thermal states in absence of any other physical 
 requirements as, for example, some regularity prescription for 
 the renormalized
 stress-tensor. \\
 A second method, which from now on  will be called 
 the HNS principle, was introduced by
 Haag, Narnhofer and Stein in 1984
 \cite{haag} (see also \cite{FH,libro haag}) and successively developed by 
  Hessling \cite{hessling}. 
 This method is connected to the well known 
 Hadamard expansion of two-point Green functions in a curved space-time
 in the limit of  the coincidence of the arguments.
  Haag, Narnhofer and Stein in \cite{haag}
 proved that if one assumes fairly standard axioms of
 quantum (quasi-free) field theory, particularly {\em local definiteness} and 
 {\em local stability} in an at least stationary,
 causally complete space-time region,
 then, roughly speaking, the thermal Wightman functions 
  in the  {\em interior} of this region will transform  into {\em non thermal}
 and {\em massless} Wightman functions in the flat space-time 
 when the 
 ``distance'' of the arguments is  vanishing (see the formula below).\\
We will call this statement, which from a naive point of view seems
 to follow from the Einstein equivalence principle, the HNS {\em theorem}.\\ 
 The   statement above is valid also in the case $T=0$. Furthermore, one
 should note that within the framework of the HNS theorem 
  the Wightman  functions
 are properly 
 considered to be {\em distributions}. 
 For example, in case of two-point thermal Wightman functions of a 
 scalar field  it holds:
 \beq
  {\lambda}^2 W^{\pm}_{\be}(x+\lambda z_{1},x+\lambda z_{2}) \rightarrow 
\frac{1}{4\pi^{2}} 
\frac{1}{g_{\mu \nu}(x)z^{\mu}z^{\nu}} 
   \;\;\; \mbox{as} \;\;\; \lambda \rightarrow 0^{+}
 \:, \label{HNS interno} 
\eeq 
where $z=z_{2}-z_{1}$.
In the equation above the coordinates $x \equiv x^{\mu} $ 
indicate a point in the {\em interior}
of the region, 
\beq
z_{(j)} \equiv z_{(j)}^{\mu}
\frac{\partial \;\;}{\partial x^{\mu}}\mid_{x} \nn
\eeq
 indicates  vectors
 in the tangent space at $x$ 
  and finally  we used the obvious
 notation: 
\beq
x+z_{(j)} \equiv x^{\mu}+z_{(j)}^{\mu}\:.\nonumber
\eeq  
Both sides of Eq.(1) are  distributions acting on 
a couple of smooth test functions in the corresponding variables 
$z_{1}$ and $z_{2}$.\\
 Finally  Haag, Narnhofer and Stein {\em principle} generalises
 HNS theorem and it affirms 
 that in the case the  space-time region we are dealing with
 is 
 just {\em a part} 
 of the whole
 manifold separated by {\em event horizons}, 
 the  point coincidence behaviour 
 of the  Wightman functions for a physically
 sensible (~thermal or not ) state must hold also 
{\em onto the  horizons}.\\
 Haag Narnhofer and Stein proved 
 in \cite{haag} that in the case of Rindler and Schwarzschild space-times,
 this constraint holds   only for $\be_{T}= \be_{M}$, the same value 
 obtained by the first method.   
 The HNS  principle determines the Unruh and  Hawking temperatures.\\
 Actually, the above statement  
 holds even if  a weaker version of HNS principle is used.
 Indeed one must be careful in using literally Eq.(\ref{HNS interno}) because 
 therein  the time
 component of the vector $z$ is endowed with a small imaginary part $\mp i\ep$
 and it is understood the $\ep-$ prescription which 
 involves
 two  weak limits as $\ep\rightarrow 0^{+}$ first and
 $\lambda \rightarrow 0^{+}$ afterward.
 So Haag Narnhofer an Stein, on their way to find the Hawking 
 temperature, interpreted the Wightman functions in 
 Eq.~(\ref{HNS interno}) 
 strictly as  functions. In this sense Eq.~(\ref{HNS interno}) 
 implies:
 \beq
  {\lambda}^2 W^{\pm}_{\be}(x,x+\lambda z) \rightarrow \frac{1}{4\pi^{2}}
\frac{1}{g_{\mu \nu}(x)z^{\mu}z^{\nu}}
   \;\;\; \mbox{as} \;\;\; \lambda \rightarrow 0^{+} \:, \label{HNS interno2}
\eeq
  In fact it can be simply  proved that the stronger
  version of the HNS principle implies the weaker one, formally
 expressed by Eq.~(2), by using smooth test functions $f(z_{1})$ and $f(z_{2})$
 in (1) which are not ``light-like correlated'', 
 i.e., such that 
  all the vectors $z=z_{2}-z_{1}$ are not {\em light-like} whenever 
 $f(z_{1})\neq 0$ and $f(z_{2})\neq 0$.
  Generally speaking, this interpretation 
 does not  eliminate the $\ep-$prescription, as it is
  also necessary to deal with
 possible cuts in the complex time plane, 
 but it transforms the weak limits
 in Eq.~(\ref{HNS interno}) into  usual limits of functions.\\
 Following the original paper of Haag Narnhofer and Stein we interpret 
 Eq.~(\ref{HNS interno})
 in this weaker sense  and thus we 
 evaluate the limit in Eq.~(\ref{HNS interno2}) 
 for {\em real} vectors $z$, space-like or
 time-like; for light-like vectors we expect a divergent 
 limit\footnote{In our massless case this is equivalent to work with
 the Hadamard function instead of the Wightman functions, see below.}.\\
 In order to  use the (weak) HNS principle for two-point Wightman functions
 one has
 to check  their behaviour as one point is fixed on the horizon and the 
 other is running toward the first from the inside of the considered region.
 This is not as simple as one might think at first, because 
 the  metric could become degenerated  on the horizon in the stationary 
 coordinates which
 define the studied thermal Wightman functions, consequently it
 could not be possible to write down the right hand side of Eq.~(2) in that
 coordinate frame.
 However, as pointed by Haag, Narnhofer and Stein in \cite{haag},
 one can check the 
 validity of the HNS principle in stationary coordinates 
 using directly Eq.~(2)  for  specific points on the  horizons, i.e.,
 for those which
 belong to the intersection of the past and the future horizons, and along 
 appropriate directions,
 but not on the whole horizon. It is very interesting to note that the check 
 of the behaviour of Wightman functions in  these
 ``few'' points  is sufficient to determine the Unruh and the 
 Hawking temperatures respectively in the case of the Rindler wedge
 and the Schwarzschild background. 
 Hessling proved in \cite{hessling} that the HNS principle, in the case of 
 the Rindler wedge, determines
 the Unruh temperature only when it works on the intersection of the horizons,
 but it does not determine any temperature by 
 considering the remaining points. In {\bf Section 1} we will report an 
 independent proof of this fact.\\ 
 Unfortunately in  the case of an extremal Reissner-Nordstr\"{o}m black hole
 the past and the future horizons  do not intersect and furthermore
 it is not possible to deal with the original procedure of Haag, Narnhofer 
 and Stein because in this case some important technical 
 hypothesis does not hold, 
 e.g., the requirement  of a 
 non-vanishing surface gravity or related parameters
 \cite{haag}, consequently, up to now, there is no proof of the 
 validity of the  HNS principle for some (or every!) value of $\be$ in 
 this case.\\
 In this situation the Hessling development of the HNS principle
 results to be
 very useful because, in the Minkowski space-time at least,
 it determines the Unruh temperature  working also
 considering  the points which do not belong to the intersection of the 
horizons
 \cite{hessling}.  In {\bf Section 1} we  will  report an independent
 proof of this fact, too.
 We may expect a similar result in the case of an extremal  R-N black hole
 where the intersection does not exist.\\   
 Furthermore \cite{hessling} one could note that, in a 
 real (Schwarzschild)  black hole, the {\em past} event
 horizon does not exist and thus also the  intersection
 of horizons does not exist. For this reason the Hessling principle
 result to be 
 very important as far as the possibility to use this in more physical
 situations than the  eternal black hole cases is concerned.\\ 
 {\em Hessling's principle}, trying to define a
{\em  Quantum (Einstein) Equivalence Principle} to be imposed on the
 physically sensible quantum states 
 \cite{hessling}\footnote{Really, Hessling considers 
 in \cite{hessling} also $n$-point functions, but we will restrict
 our discussion by considering only the case of a scalar quasifree
 field and thus by studying only the two-point functions.}, 
 requires the {\em existence} of the limit:
 \beq
 \lim_{\lambda\rightarrow 0^{+}}
 N(\lambda)^{2}\: W^{\pm}_{\be}(x+\lambda z_{1},x+\lambda z_{2}) \:,
 \label{hessling1} 
 \eeq
as a {\em continuous} function of $x$,
 for some function $N(\lambda)$ monotonous and nonnegative for $\lambda>0$. 
 Furthermore, it requires the validity of a much more strong condition
 in every {\em local inertial coordinate system} around
 $x$ (i.e., a coordinate frame such that
 $g_{\mu\nu}(x)=diag(-1,1,1,1)$, $\partial_{\rho} g_{\mu\nu}(x)=0$): 
 \beq
 \lim_{\lambda\rightarrow 0^{+}}\frac{d\:\:}{d\lambda} 
 \: N(\lambda)^{2} W^{\pm}_{\be}(x+\lambda z_{1},x+\lambda z_{2}) = 0\:.
 \label{hessling2}
 \eeq
 The latter requirement
 is connected with the fact that within a local inertial
 system the metric looks like the Minkowski metric {\em up to the first order}
 in the coordinate derivatives. In the Minkowski
 background and using Minkowskian coordinates 
 Eq.(\ref{hessling2}), which involves coordinate derivatives
{\em up to the first order}, results to be satisfyed and thus we expect this
will hold in curved backgrounds by using local inertial coordinate 
systems\footnote{We stress that Hessling generalises the above
  equations in order to be able to use  a  non local inertial coordinate 
  system \cite{hessling}, too.},
too (see \cite{hessling} for details).
We can note that, away from the 
horizons, the validity HNS {\em theorem} implies the validity
of the  first
 Hessling requirement with $N(\lambda)=\lambda $.
Furthermore, the validity of the HNS principle on a horizon implies
the validity of the first Hessling requirement there.\\
As in the case of the HNS principle, 
we will use the Hessling principle on an event horizon in a {\em weaker}
 version, 
by checking the validity
of Eq.(2) as well as of the following equation:
\beq
\frac{d\:\:}{d\lambda}
{\lambda}^2 W^{\pm}_{\be}(x,x+\lambda z) \rightarrow 0
   \;\;\; \mbox{as} \;\;\; \lambda \rightarrow 0^{+} \:, \label{hessling}
\eeq
where $x$ belongs to the horizon, $z$ is a {\em real space-like or time-like}
 vector  and a  {\em local inertial
coordinate system} is used.\\
It is interesting to check whether 
 the HNS principle and the Hessling principle select special
 temperatures or, like the  method mentioned first,
 accept every temperature for thermal states in the  case of
 an extremal Reissner-Nordstr\"{o}m black hole. This question arises also 
 because in the case of the Rindler and Schwarzschild space-times
 the use of the HNS and Hessling principle seems to be  more selective
 than the usually used  first method.
 Indeed,  employing the   HNS (or the Hessling) principle, one  obtains
 the result that
 all the temperatures of the form $T_{n}=n/\be_{M}$ with $n=2,3,...$ must 
 be rejected which 
 otherwise would be
 permitted\footnote{Actually, as for example in the case of the Rindler space,
  other physical requirements (e.g.,the behaviour of the
  stress-tensor on horizon) reject these temperature too.}.\\
 Finally, it seems  necessary to spend some words about the 
very important {\em Kay-Wald theorem} \cite{wald}. They proved that in a
 space-time with a 
{\em bifurcate Killing horizon} \cite{wald} (furthermore endowed with
 an opportune discrete ``wedge reflection'' isometry) like the Minkowski
 manifold
endowed with Rindler's wedges as well as the Kruskal space-time endowed with
Schwarzschild wedges,  every  quasifree {\em stationary}
 state (with respect to 
the Killing vector defining the horizons) satisfying the 
{\em Hadamard condition}\cite{wald,libro haag} 
in a neighborhood of the horizon, results to be a
 KMS state inside of the wedges where the Killing vector 
is time-like. Furthermore the temperature results to be 
  the Unruh-Hawking temperature, too.\\ 
  We want to stress  that in  the background of an extremal
  Reissner-Nordstr\"{o}m black hole it is impossible to use the Kay-Wald 
 theorem because it needs explicitly a {\em not empty}
 intersection of horizons 
 \cite{wald}. \\

 In {\bf Section 1} we will
 start with some aspects of the well known 
 Rindler theory
 to  observe some interesting  features of the HNS principle and 
 the Hessling principle by extending the 
 ``horizon-check'' onto  the whole horizon recovering the same
results found by Hessling \cite{hessling}.\\
 In {\bf Section 2}, by using this generalised method, we will
 study  
 the temperatures of thermal states in the case of a extremal
  Reissner-Nordstr\"{o}m 
 black hole for a scalar field and in the limit of a 
 black hole with large mass.  We will take advantage there of the vanishing
 scalar
 curvature of the manifold and  we will use the resulting coincidence
 of the  conformal coupling
 with the minimal coupling for a massless scalar field.\\
 Finally, in {\bf Section 3} we will discuss our results.


\section{HNS and Hessling's principles in the Rindler Space }

  We consider the Minkowski space-time with signature:
\beq
 g_{\mu\nu}\equiv \mbox{diag}(-1,1,1,1)\:. \nonumber
\eeq
 In this manifold we consider a local coordinate frame
 $(\rho, \tau, x^{\perp})$ 
  connected to Minkowski coordinates $(x^0,x^1,x^2,x^3) = (t, \vec{x})$ 
  by the equations:
 \beq
  x^0 = \rho \sinh \tau \:,
  \eeq  
 \beq
 x^1 = \rho \cosh \tau \:,
 \eeq
 \beq  
 x^{\perp} = (x^2, x^3) \:,
 \eeq
 where $ \tau, x^2, x^3  \in ( - \infty , + \infty )$ and $\rho > 0 $.\\
 This region is called {\em Rindler wedge} \cite{rindler}. 
 In Minkowski coordinates it is:
 \beq
 x^1>|x^0| \:.  \nonumber
 \eeq
 In Rindler coordinates, the Minkowski metric is written as:
 \beq
 ds^2=-\rho^{2}d\tau^{2}+d\rho^{2}+(dx^{\perp})^{2}
 \:. \label{metrica Rindler}
 \eeq 
 Note that the part:
  \beq 
H^{+} = \{ x^1=x^0 , x^1 \geq 0 \} 
 \eeq
  of 
 the boundary
 of the Rindler wedge is the {\em future}
 event horizon  and the part:  
 \beq
H^{-} = \{ x^1= -x^0 , x^1 \leq 0 \} 
\eeq 
 is the {\em past} event horizon 
 of the region.\\
 
 Now we calculate the  Wightman functions (see \cite{fulling} for example)
 and the  thermal
 Wightman functions (see \cite{rindler} for example) of a massless field
 and then discuss the HNS principle in this simple 
 case.\\
 To construct the canonical theory \cite{fulling} of a massless scalar
 field propagating in the Rindler wedge we expand the field operator into 
 positive and 
 negative
 frequency modes of the time-like Killing vector field tangent to
 the $\tau$ coordinate.
 This vector field also generates  $\tau$-translations.\\
  By this way we obtain:
\beq
\phi=\int\frac{d^2k}{2\pi}\int_0^{\ii}\frac{d\om}{\pi}[\sinh(\pi\om)]^{1/2}
K_{i\om}(k\rho)\left[A_{k\om}
e^{i(kx^{\perp}-\om\tau)}+A^{\dagger}_{k\om}e^{-i(kx^{\perp}-\om
\tau)}\right] \:.
\eeq
 Then we impose the commutation relations which are
  equivalent to the usual canonical 
 commutation rules:
\beq
\left[A_{k\om}, A_{k^{'}\om^{'}}\right]= 0 \:,\nonumber
\eeq
\beq
\left[A_{k\om},A_{k^{'}\om^{'}}^{\dagger}\right]=\de(k-k^{'})\de(\om-\om^{'})
 \:, 
\eeq
\beq
\left[A_{k\om}^{\dagger},A_{k^{'}\om^{'}}^{\dagger}\right]= 0 \:. \nonumber
\eeq
 It is easy to see that this fact follows from the orthogonality and
 completeness relations:
\beq
\int_0^{\ii} K_{i\om}(k\rho)K_{i\om^{'}}(k\rho)\rho^{-1}d\rho=
\frac{\pi^2}{2\om
\sinh\pi\om}\de(\om-\om^{'}) \:,
\eeq
\beq
\int_0^{\ii}\frac{2\om\sinh\pi\om}{\pi^2}K_{i\om}(k\rho)K_{i\om}(k\rho^{'})
d\om
=
\rho\de(\rho-\rho^{'}) \:.
\eeq
$K_{i\om}(k\rho)$ denotes the Mc Donald functions of
imaginary order $i\om$.\\
 We obtain the Wightman elementary functions by calculating 
the  bra-ket average
   of a two field operator product in the 
 vacuum state, annihilated by $A_{k\om}$ \cite{fulling}. They are:
\beq
W^{+}(x,x^{'})=\int\frac{d^2k}{4\pi^2}e^{ik(x^{\perp}-x^{'\perp})}\int_0^{\ii}
\frac{d\om}{\pi^2}\sinh\pi\om
e^{-i\om(\tau-\tau^{'}-i\ep)}K_{i\om}(k\rho)
K_{i\om}(k\rho^{'}) \:, 
\eeq
\beq
W^{-}(x,x^{'})=\int\frac{d^2k}{4\pi^2}e^{ik(x^{\perp}-x^{'\perp})}\int_0^{\ii}
\frac{d\om}{\pi^2}\sinh\pi\om
e^{i\om(\tau-\tau^{'}+i\ep)}K_{i\om}(k\rho)
K_{i\om}(k\rho^{'}) \:,
\eeq
where we used the abbreviation $x=(\tau,\rho,x^{\perp})$.\\ 
It is well known that the other Green functions 
can be built up with the Wightman elementary functions \cite{fulling}.\\
We can integrate the expression for $W^{\pm}(x,x^{'})$ and 
obtain \cite{dowker3,dowker4}:
\beq
W^{\pm}(x,x^{'})=-\frac{1}{4\pi^2}
\frac{\al}{\rho\rho^{'}\sinh\al}\frac{1}{(\tau-\tau^{'}\mp i\ep)^2-\al^2}\:,
\label{nontermica}
\eeq
where:
\beq
y =\frac{\rho^2+\rho^{'2}+|x^{\perp}-x^{'\perp}|^2}{2\rho\rho^{'}} 
\; \; \; \mbox{and} \; \; \;
\cosh \al = y  \label{alpha}\:.
\eeq
Note that $W^{\pm}$ are  distributions.\\ 
It is  important to stress that, for $\ep=0$ and
 $\rho,\rho^{'},x^{\perp},x^{'\perp}$ fixed, the 
 right hand side of Eq.~(\ref{nontermica}),  strictly considered as a 
  function, can be analytically
extended  to a function $W$ defined in the whole 
 complex $(\tau-\tau^{'})$-plane except at the poles 
$(\tau-\tau^{'}) = \pm \alpha$. It is also important to note 
that we would have obtained
exactly the same extension by starting from $W^{-}$ or $W^{+}$ 
because of the triviality of
equal-time commutation relations.
In terms of $W$ Eq.~(\ref{nontermica}) reads:
\beq
W^{\pm}(\tau-\tau^{'})=W(\tau-\tau^{'}\mp i\ep)\:, \nonumber
\eeq
where $\tau-\tau^{'}$ is  real valued now.
The $\ep$-prescription indicates the manner in which 
 one obtains distributions
 $W^{+}$ 
and $W^{-}$ as weak limits.
According to the KMS condition for bosons \cite{kms},
 this extended thermal Wightman 
 function $W_{\be}$ must be
periodic in imaginary time with period $\be$. 
One can find $W_{\be}$  with the {\em method of images} \cite{fr,rindler}.
It reads:
\beq 
W_{\be}(x,x^{'})=-\frac{1}{4\pi^2}\frac{\al}{\rho\rho^{'}\sinh\al}
\sum_{n=-\ii}^{\ii}\frac{1}{(\tau-\tau^{'}-in\be)^2-\al^2} \:.\nonumber   
\eeq
 We can sum the series, the result being:
\beq
W_{\be}(x,x^{'})=\frac{1}{4\pi^2}\frac{\pi\left\{\coth\frac{\pi}{\be}
(\al+\tau-\tau^{'})+\coth\frac{\pi}
{\be}(\al-\tau+\tau^{'})\right\} }{2\be\rho\rho^{'}\sinh\al}\:.\nonumber
\eeq
To make it clear, this expression indicates the whole complex extension of
 {\em both}
of the thermal Wightman functions.
In our  massless case this function with  $\tau-\tau^{'}$
  real valued coincides with the {\em Hadamard function} $W^{(1)}:=
W^{+}+W^{-}$ \cite{fulling,rindler} except for a factor $1/2$.
 Finally, one return to the real time
 thermal Wightman functions in the { \em distributional} sense 
by restoring the usual 
$\ep$-prescription obtaining \cite{dowker4}:
\beq
W^{\pm}_{\be}(x,x^{'})=\frac{1}{4\pi^2}
\frac{\pi\left\{\coth\frac{\pi}{\be}(\al+\tau-\
\tau^{'}\mp i\ep)+\coth\frac{\pi}
{\be}(\al-\tau+\tau^{'}\pm i\ep)\right\} }{2\be\rho\rho^{'}\sinh\al}\:.
\label{termica}
\eeq 
In this {\em massless} case the singularities 
at $\tau-\tau^{'}=\pm\alpha$ are poles and therefore
 it is not necessary to use 
the $\ep-$prescription for the  functions,
but if we deal with massive fields 
these poles become branching  points \cite{fr}.
In this situation the $\ep-$prescription, in case of causally related
 arguments  $x$ and $x^{'}$, tells us how to calculate the
 limit, depending on the side from which we approach the cuts, in order to
 distinguish 
 $W_{\be}^{-}$ from $W_{\be}^{+}$.\\

In order to find  the Unruh temperature we should note that the metric 
could become singular
if we extend $\tau$ to imaginary values: $\tau \rightarrow -i\tau$
 \beq
ds^2=\rho^{2}d\tau^{2}+d\rho^{2}+(dx^{\perp})^{2}\:. 
\label{metrica Rind euclidea}
 \eeq
Indeed, let $\be$ be the period of the imaginary time coordinate.
 If $\be \neq 2\pi $
then the metric  will have a  non trivial 
 {\em conical}-like singularity at $\rho=0$ (however it is possible to
study the quantum field theory also in this background
\cite{conic0,conic1,conic2,conic3}).
Thus we see that
 $\be=2\pi$ is the only choice in order to have 
a  globally regular manifold\footnote{This requirement arises when
one tries to use  the
 functional integral over the gravitational
configurations as well as over
 the quantum field configurations \cite{conic-1}.}.
This fact implies that the period of the thermal green functions must be  of
 the kind:  
\beq
\be_{k} = \frac{2 \pi }{ k} \:,    \label{periodi}
\eeq
where $k=1,2,3 ...$.
It means that {\em a priori} the possible temperatures are just of the form:
\beq
T_{k} = \frac{k}{2\pi}     \:,
\eeq 
where $k=1,2,3 ...$\\
The value: $T_{U}=T_{1}=\frac{1}{2\pi}$ is the well known 
{\em Unruh temperature} 
(for complete references see: \cite{giapponese,libro haag,rindler}).\\
 After some algebra with the hyperbolic functions, we
 recover the well known result for  $\be_{U}=\be_{1}=2\pi$, holding
inside of a Rindler wedge and in the sense of the distributions:
\begin{eqnarray}
W^{\pm}_{2\pi}(x,x^{'}) &=&\frac{1}{4\pi^2}\frac{1}{\rho^2+\rho^{'2}+
|x^{\perp}-x^{'\perp}|^2-2\rho\rho^{'}
\cosh(\tau-\tau^{'}\mp i\ep)} \nn \\
&=&\frac{1}{4\pi^2}\frac{1}{-(t-t^{'}\mp i\ep)^2+(x^1-x^{'1})^2+|x^{\perp}
-x^{'\perp}|^2}\nn \\
&=&\frac{1}{4\pi^2}\frac{1}{\si^2(x_{\mp \ep},x^{'})}\:,
\end{eqnarray}
where we used the notation  $x_{\mp \ep}\equiv (t\mp i\ep,\vec{x})$ and 
 introduced the geodesic invariant distance:
\begin{eqnarray}
\si^2(x,y)&=&-(t_{x}-t_{y})^2+(x_{x}^1-x_{y}^{1})^2+|x_{x}^{\perp}
-x_{y}^{\perp}|^2\\
&=&-(t_{x}-t_{y})^2+|\vec{x}_{x}-\vec{x}_{y}|^2\:.
\end{eqnarray}\\
$W^+_{2\pi}$ is exactly the Wightman function  of the Minkowski
vacuum (thus it satisfies the HNS and Hessling prescriptions
 everywhere trivially).
This means that the Minkowski vacuum is a KMS state with respect to
$\tau$-translations.
 There exists a vast literature 
about this topic; for a 
complete review see \cite{libro haag} and \cite{giapponese}.\\

Now we will check the HNS and the Hessling principle on the whole 
horizon for arbitrary values of $\be$.
Incidentally, it shall be mentioned that  one can check the correct behaviour 
of Wightman functions away from  the  horizons with no particular 
difficulties but in this paper we will
deal with the   Wightman functions behaviour  on the horizons
exclusively.\\
Note that in Rindler coordinates the  coordinate representation of the metric
(but not the metric) becomes degenerated onto the horizons, so
 there seems to be the necessity to introduce new,
generally non stationary,  coordinate frames
in a neighborhood of every point of the horizons, to check the HNS and
 Hessling principles.
The price that must be payed  when one uses
  non stationary coordinates is that 
 one has to drop
the translational time-invariance of the Wightman (and Green) functions
and thus the theory becomes more complicated.  
(~Actually, as pointed out in the introduction, the use of  new coordinates 
 is not necessary for the points 
 which belong to 
 $H^{+} \cap H^{-}$ \cite{haag}, but in our aim to be more general, we want
to deal with the whole horizon).\\
In the Rindler space-time one could use Minkowski coordinates, but in other 
  curved space-times  it is rather unlikely to find  similar
 simple
 coordinate frames! Thus we will  
develop a method which employs only  stationary
coordinates wherein 
 the thermal theory
is much more simple. \\
Indeed it is possible to check the HNS and the Hessling principles by studying 
the behaviour of two point Wightman functions along every geodesic which starts
from the event horizons.  We  note that for every geodesic which meets
the horizon in a point $x$ there is a locally geodesic coordinate frame 
with
the origin at  $x$. Furthermore we can 
chose the  geodesic to be a  coordinate
axis, the running coordinate being just the length of the  geodesic measured 
from $x$.
By varying the  geodesics which meet the horizon one obtains all the possible
points of the horizon together with  their tangent vectors.
 We stress that it is possible to execute this checking procedure also 
by using  stationary 
coordinates
(Rindler coordinates in this case), because the geodesic length is invariant
and so is not sensitive to the degeneracy in this representation of the
metric. \\
Let us illustrate this method for the Rindler case.\\

For sake of simplicity we will only  deal with geodesics in a plane 
$x^{\perp}= $constant.
 For
geodesics which meet the horizon we obtain  some useful formulas  
by translating the linear geodesic
equations from  Minkowski coordinates into  Rindler coordinates.\\

{\em Geodesics which meet} $H^{+}\cap H^{-}$.
\begin{eqnarray}
 \tau&=&\tanh^{-1}\alpha    \label{uno}\\
\rho&=& s \label{due}\\ 
s&=&x^{1} \sqrt{1-\alpha^{2}}\:, \label{tre} \\
\end{eqnarray} 
where $\alpha \in (-1,+1)$ is a constant parameter and  $s$ is the geodesic 
length
measured from the horizon.  It can be easily proved that all
 these geodesics are space-like.\\

{\em Space-like geodesics which meet} $H^{+}-(H^{+}\cap H^{-})$.
\begin{eqnarray}
 \tau&=&\gamma +
 \coth^{-1}\left(1+\frac{se^{\gamma}}{y}\right)\:  \left( = 
\gamma+\sinh^{-1}\left(\frac{ye^{-\gamma}}{\rho}\right)
 \right)  \label{quattro}\\
\rho&=& ye^{-\gamma}
\sqrt{\left(1+\frac{se^{\gamma}}{y}\right)^{2}-1} \:
 \left( = \frac{ye^{-\gamma}}{\sinh(\tau-\gamma)} \right) \label{cinque} \\
s&=&ye^{-\gamma}\left(\coth(\tau-\gamma)-1\right)\:, \label{sei}
\end{eqnarray}
where $\gamma \in (-\infty,+\infty)$ is a constant parameter,
  $s$ is the geodesic length measured from the horizon and $y= x^{1}= x^{0}$ 
 is the coordinate of the intersection of the geodesic and
 $H^{+}$.\\

{\em Time-like geodesics which meet} $H^{+}-(H^{+}\cap H^{-})$.
\begin{eqnarray}
 \tau&=&\gamma + 
\tanh^{-1}\left(1-\frac{se^{\gamma}}{y}\right)
\:\left( = \gamma+\cosh^{-1}\left(\frac{ye^{-\gamma}}{\rho}\right)
 \right)   \label{sette}\\
\rho&=& ye^{-\gamma}
\sqrt{1-\left(1-\frac{se^{\gamma}}{y}\right)^{2}}
 \:  \left( =  \frac{ye^{-\gamma}}{\cosh(\tau-\gamma)} \right) 
\label{otto} \\
s&=&ye^{-\gamma}\left(1-\tanh(\tau-\gamma)\right) \:,\label{nove}
\end{eqnarray}
where $\gamma \in (-\infty,+\infty)$ is a constant parameter,
$s$ is the geodesic length measured from the horizon and $y= x^{1}= x^{0}$
 is the coordinate of the intersection between the geodesic and
 $H^{+}$.\\

 One can find similar formulas for geodesics which meet $H^{-}$, but we
deal only with geodesics falling into $H^{+}$ because of
 the trivial time-symmetry of the problem.
In fact the Minkowski time-reversal transformation:
\beq
t \rightarrow -t  \nonumber 
\eeq
\beq
\vec{x}\rightarrow \vec{x} \nonumber
\eeq
is equivalent to the Rindler time-reversal transformation:
\beq
\tau \rightarrow -\tau  \nonumber 
\eeq
\beq
\rho \rightarrow \rho \nonumber
\eeq
\beq
x^{\perp} \rightarrow x^{\perp} \nonumber
\eeq
One obtains the geodesics meeting $H^{-}$ by using the above time 
transformation into geodesics which meet $H^{+}$. Furthermore the 
same time-reversal action transforms the Wightman functions of 
Eq.~(\ref{termica})
into their complex conjugate. In this way can be easy proved that
the Wightman functions will satisfy HNS and Hessling's 
prescription on $H^{-}$ if they
do so on $H^{+}$.\\ 

We start studying the geodesics which run through the origin of 
Minkowski coordinates. 
We obtain, by inserting (\ref{uno}) and (\ref{due}) into (\ref{alpha}) and 
 (\ref{termica}) and by calculating  the limit as 
$s=\rho \rightarrow 0$:
\beq
W^{\pm}_{\be}(x_{H^{+}},x^{'})=\frac{1}{4\pi^{2}} 
\frac{2\pi}{\be \rho^{'2}} \:,\nonumber
\eeq 
where $x_{H^{+}}\equiv (x^{0}=0,x^{1}=0,x^{2},x^{3})$.\\
Using the same notations as in Eq.~(2), we substitute $\rho^{'}=s$ 
 thinking $s\equiv z^{i}$ to be  
an increase 
of a space-like 
coordinate of a coordinate frame in  $x_{H^{+}}$,
as well as to be the only non-vanishing component of a 
tangent vector $z$ in $x_{H^{+}}$.
The result reads:
\beq
\lambda^{2} W^{\pm}_{\be}(x_{H^{+}},x_{H^{+}}+\lambda z)=
\lambda^{2} W_{\be}(x_{H^{+}},x_{H^{+}}+\lambda z) =    \nn
\eeq
\beq
=\lambda^{2}\frac{1}{4\pi^{2}}
\frac{2\pi}{\be \lambda^{2} z^{i2}} = 
\frac{1}{4\pi^{2}}\frac{2\pi}{\be} 
\frac{1}{g_{\mu\nu}(x_{H^{+}})z^{\mu}z^{\nu}}\:.\nonumber
\eeq
Thus we see that, in order to be consistent with
 the HNS principle, $\be=2\pi$ is the only possible choice.
Before to consider the other cases we note that if we tried
 to repeat  the
calculations above with the {\em zero temperature} green function 
(\ref{nontermica}),
 we would fall in trouble.
In fact, as $x \rightarrow H^{+}\cap H^{-}$ along the considered 
geodesics:
\beq
W^{\pm}(x,x^{'}) \rightarrow 0 \:, \nonumber
\eeq
for every $x^{'}$.
We conclude that the requirement of the validity of the  HNS principle rejects 
also the {\em Rindler vacuum} which  for that can not be considered as a 
 physically sensible state for the  whole Minkowski 
space-time\footnote{
If we consider rather the Rindler wedge as the whole physical manifold,
then we can consider the Rindler vacuum as a physical state. Note that in 
such a  situation the prescriptions for renormalizing  physical 
quantities also change, because it is not opportune to  subtract
 the corresponding Minkowski quantities.}. 
Note that the Rindler vacuum
is outside of the  Fock representation of Hilbert space generated by
Minkowski vacuum and {\em vice versa}. Indeed  it is well known that
the Minkowski vacuum 
requires a vanishing normalization coefficient
when it is built up  by the normal modes of the Fock representation 
generated by
the (left hand and right hand) Rindler vacuum \cite{giapponese}.\\
 However within the framework of the usual {\em algebraic approach}
to quantum field theory \cite{libro haag,wald} 
such a situation is quite common 
 and, differently from the HNS prescription, it 
 does not distinguish directly between physical and
unphysical states. \\
 
Now we consider the case of the space-like geodesics which fall 
into $H^{+}$, in a point $x_{H^{+}}\equiv (y,y,y^{\perp})$.\\
We obtain by inserting (\ref{quattro}) and (\ref{cinque}) into 
(\ref{alpha}) and
 (\ref{termica}) and by calculating  the limit as
$s=\rho \rightarrow 0$:
\beq
  W^{\pm}_{\be}(x_{H^{+}},x^{'}) =  
\frac{1}{4\pi^{2}}
     \frac{\pi}{\be \rho^{'2}}
 \left\{ 1+ \coth \left[ \frac{\pi}{\be} \left( 
 \tau^{'} + \ln \frac{\rho^{'}}{2y}
 \right) \right]   \right\} \:.\label{fondamentale}
\eeq
It is important to stress that this function  diverges for: 
\beq
\tau^{'} + \ln \frac{\rho^{'}}{2y}=0\:, \nonumber
\eeq
 i.e., on {\em null} geodesics
which fall in $x_{H^{+}}$, just   as we have expected.\\
Let us examine the Wightman function behaviour on space-like geodesics.\\
 Using again (\ref{quattro}) and (\ref{cinque}) 
for the variables $\tau^{'}$ and $\rho^{'}$, substituting
 $s=\lambda z^{i}$ and finally  calculating the limit as 
$\lambda \rightarrow 0$ we obtain $(W^{\pm}=W)$:
\begin{eqnarray}
\lambda^{2} W_{\be}(x_{H^{+}},x_{H^{+}}
+\lambda z) & = & \frac{1}{4\pi^{2}}
\frac{\pi\lambda \left\{ 1+ \coth\left[\frac{\pi}{\be}\ln \left(1+
\frac{\lambda z^{i} e^{\gamma}}{2y}
 \right)\right]\right\} }{\be e^{-\gamma}y z^{i} \left(2+\frac{\lambda z^{i}
 e^{\gamma}}{y}
 \right)}  \label{first} \\
&\sim& \frac{1}{4\pi^{2}} 
  \frac{\pi}{2\be e^{-\gamma}y z^{i}}
\frac{\lambda}{ \sinh \left( \frac{\pi}{\be} 
\ln \left( 1+
\frac{\lambda z^{i} e^{\gamma}}{2y} \right) \right)  } \nonumber\\
&\sim& \frac{1}{4\pi^{2}}
  \frac{\pi}{2\be e^{-\gamma}y z^{i}}
\frac{\lambda}{  \frac{\pi}{\be}
\ln \left( 1+
\frac{\lambda z^{i} e^{\gamma}}{2y} \right)  } \nonumber\\
&\sim& \frac{1}{4\pi^{2}}
  \frac{1}{2 e^{-\gamma}y z^{i}}
\frac{\lambda}{ 
\frac{\lambda z^{i} e^{\gamma}}{2y}  } \nonumber\\
&=& \frac{1}{4\pi^{2}}\frac{1}{z^{i2}} =
  \frac{1}{4\pi^{2}} \frac{1}{g_{\mu\nu}(x_{H^{+}})z^{\mu}z^{\nu}}
\:.\nonumber
\end{eqnarray}
 We found that the 
HNS principle holds for this kind of geodesics for all values  $\be>0$.\\

Finally we consider the case of time-like geodesics meeting $H^{+}$.\\
After some limit calculations  as
$s \rightarrow 0$ in the first variable $x$ of Wigthman function
(\ref{termica})
we obtain, inserting (\ref{sette}) and (\ref{otto}) into (\ref{alpha})
and (\ref{termica}):
\beq
 W^{\pm}_{\be}(x_{H^{+}},x^{'}) =
\frac{1}{4\pi^{2}}
     \frac{\pi}{\be \rho^{'2}}
 \left\{ 1+ \coth \left[ \frac{\pi}{\be} \left(
 \tau^{'} + \ln \frac{\rho^{'}}{2y}
 \right) \right]   \right\} \nonumber
\eeq 
 This expression for  $W^{\pm}_{\be}$ coincides with the right hand side
 of  Eq.~(\ref{fondamentale}).
Indeed the value of a Wightman function with one point on the horizon must not 
depend on
the direction from which we reach the horizon!
Now,  by using again (\ref{sette}) and (\ref{otto})
for the variables $\tau^{'}$ and $\rho^{'}$, by substituting
  $s=\lambda z^{0}$ with
$\lambda \rightarrow 0$ we obtain ($W^{\pm}=W$):
\begin{eqnarray}
\lambda^{2} W_{\be}(x_{H^{+}},x_{H^{+}}
+\lambda z) & = & \frac{1}{4\pi^{2}}
\frac{\pi\lambda \left\{ 1+ \coth\left[\frac{\pi}{\be}
\ln \left(1-\frac{\lambda z^{0} e^{\gamma}}{2y}
 \right)\right]\right\}}{\be e^{-\gamma}y z^{0} \left(2-\frac{\lambda z^{0}
 e^{\gamma}}{y}
 \right)}  \label{second} \\
&\sim& \frac{1}{4\pi^{2}}
  \frac{\pi}{2\be e^{-\gamma}y z^{0}}
\frac{\lambda}{ \sinh \left( \frac{\pi}{\be}
\ln \left( 1-
\frac{\lambda z^{0} e^{\gamma}}{2y} \right) \right)  } \nonumber\\
&\sim& \frac{1}{4\pi^{2}}
  \frac{\pi}{2\be e^{-\gamma}y z^{0}}
\frac{\lambda}{  \frac{\pi}{\be}
\ln \left( 1-
\frac{\lambda z^{0} e^{\gamma}}{2y} \right)  } \nonumber\\
&\sim& -\frac{1}{4\pi^{2}}
  \frac{1}{2 e^{-\gamma}y z^{0}}
\frac{\lambda}{ \frac{\lambda z^{0} e^{\gamma}}{2y}  } \nonumber\\
&=& -\frac{1}{4\pi^{2}}\frac{1}{z^{02}} =
  \frac{1}{4\pi^{2}} \frac{1}{g_{\mu\nu}(x_{H^{+}})z^{\mu}z^{\nu}}
\:.\nonumber
\end{eqnarray}
Once more the HNS principle  holds  for the considered
 points on the horizon  and for all the values $\be>0$.
We conclude that 
{\em only the points in} $H^{+} \cap H^{-}$ 
{\em really select a  temperature of thermal states} in the framework of the 
{\em HNS principle}.\\
 
Let us consider the {\em Hessling principle} which will produce a quite
 different result (but consistent with the previous one).\\
In the following we will define:
\beq
\mu:= Z \frac{e^{\gamma}}{2y} \:\lambda \:, \nn
\eeq
where we may understand either $Z= z^{i}$ or $Z= -z^{0}$. Furthermore, we
define:
\beq
X:= \frac{\pi}{\beta}\ln (1+\mu) \:. \nn
\eeq 
Considering Eq.(\ref{first}) as well as Eq.(\ref{second}) in order to check
Hessling's principle for points on the horizon which do not belong to
$H^{+} \cap H^{-}$, we obtain:
\beq
\frac{d\:\:}{d\lambda} \: \lambda^{2} W_{\be}(x_{H^{+}},x_{H^{+}}
 +\lambda z) = 
\frac{A}{\beta e^{2 \beta X/\pi}} 
\left\{ 1 + \coth X \: - \frac{\pi}{\beta}
\frac{ e^{ \beta X/\pi} - 1}{  \sinh^{2} X}  \right\}
\:, \nn
\eeq
where $A$ is a factor non depending on $\lambda$ and $\beta$. 
 We have to do the limit
as $\lambda \rightarrow 0^{+}$, i.e., $X \rightarrow 0$
in the right hand side of the above equation now.
Some trivial calculations lead to:
\beq
\frac{d\:\:}{d\lambda} \: \lambda^{2} W^{+}_{\be}(x_{H^{+}},x_{H^{+}}
 +\lambda z)  = 
\frac{A}{\beta} 
\: \frac{(1 - \beta/ 2\pi) X^{2}+ O(X^{3})}{X^{2}+ O(X^{3})} 
\rightarrow A \left(\frac{1}{\beta} -\frac{1}{2\pi}\right)
\:.\nn
\eeq
If we impose that the right hand side of the above equation vanishes 
as $\lambda \rightarrow 0^{+}$, i.e., $X \rightarrow 0$, 
 we will recover a Hessling result \cite{hessling}.
 Considering the points on the horizon 
which
do not belong to their intersection, 
the {\em Hessling principle} holds {\em only if} the temperature is the 
{\em Unruh temperature} ($T=1/\beta = 2\pi$)\footnote{One should observe
 that also the limit
 case $\beta \rightarrow + \infty$  does not satisfy
 the Hessling condition.}.\\
Finally, we stress that if one considers more complicated geodesics,
 i.e., geodesics 
with $x^{\perp}\neq$
 constant, all the  above results will not change, but the necessary
 calculations  are  more complicated and we will  not report on this here.\\


\section{The Case of an Extremal Reissner-Nordstr\"{o}m Black Hole}

Let us  apply the method of the first part  on the case of a massless
 scalar field in the 4-dimensional Reissner-Nordstr\"om background.
We start  with  coordinate frames (and related approximations) used in 
\cite{cvz}. \\
 The metric we are interested in reads:
\beq
ds^2=-\at1-\frac{R_H}{R}\ct^2\,(dx^0)^2+
\at1-\frac{R_H}{R}\ct^{-2}\,dR^2+R^2\,d\Omega_{2}
\:,\label{bh}
\eeq
where we are using polar coordinates, $R$ being the radial one and
$d\Omega_{2}$ being  the metric of the unit-$2$-sphere.
The horizon radius is $R_H=MG=Q$,
$M$ being the mass of the black hole, $G$ the Newton constant and $Q$
its charge.\\

It is obvious that any redefinition of  space-like coordinates does not
change the thermal properties of our field theory because these properties 
depend on the  time-like coordinate and in particular on its
tangent Killing vector. Thus, as in \cite{cvz},
 we  redefine the Reissner-Nordstr\"{o}m radial coordinate 
by means of
\beq
\rho=\frac{(\bar{R}
-1)}{1-2(\bar{R}-1)\ln(\bar{R}-1)-(\bar{R}-1)^2}
\:,\label{ro}
\eeq
where $\bar{R}=R/R_{H}$ is now implicitly defined by Eq.~(\ref{ro}).
Near the horizon $\rho \sim 0$ we can expand:
\beq
\bar{R}\sim1+\rho+O(\rho^2\ln\rho)
\:.\label{456}\eeq
In order to perform explicit computations, we shall consider
the large mass limit  of the black hole, i.e., $R_{H}\rightarrow +\infty$.
 In this limit, the whole region outside
of the black hole ($R > R_{H}$)
 tends to approach the horizon.
 Thus we can use the approximated metric near the horizon:
\beq
d s^2\sim -\rho^{2}(dx^0)^2
+\frac{1}{\rho^2} d\rho^2+ d\Omega_{2}
\label{lsus1}
\:.\eeq
We will return on this point in the final discussion.\\
Finally we change the space-like 
 frame by a new space-like coordinate:
\beq
r=\frac{1}{\rho}\:.
\eeq
The final form of the metric is very simple, it is called the 
{\em Bertotti-Robinson metric} 
\cite{bertotti robinson}:
\beq
d s^2=\frac{1}{r^{2}}\left[ -dt^2
+dr^2+r^{2}d\Omega_{2}\right]  \label{metrica}\:,
\eeq
or 
\beq
d s^2=\frac{1}{\vec{x}^{2}}\left[ -dt^2
+(d\vec{x})^{2}\right]  \label{metrica2}\:,
\eeq
where we used the obvious notation 
\beq
t=x^{0} \:,\nonumber
\eeq
\beq
\vec{x}\equiv (x^{1},x^{2},x^{3})\:. \nonumber
\eeq
Note that this metric is {\em conformal}
 to Minkowski metric by the factor
$1/r^{2}$, {\em singular} at the origin; however we are interested  in the
region near
the
horizon, i.e. $r \rightarrow +\infty$, and  this singularity is
absent there.\\
At the beginning our method involves the calculation  of 
all the possible geodesics  which start from the horizons.
It can be easily  proved that this means that we have to look for 
space-like and  time-like geodesics 
 which, starting from the  outside of the black hole
 reach the  region  $r=\infty$ in a {\em finite}
 interval of  geodesic length.
 In particular we seek the geodesics which reach  at least  one of the 
 limit  regions $r=\infty$ and $t=+\infty$
 ({\em future} horizon) or   
  $r=\infty$ and $t=-\infty$ (past horizon) in a {\em finite}
 interval of  geodesic length.\\
 Really, we will find that there exists also a kind of  
 space-like geodesics with $t=$ constant which
 just  ``seem to fall into  the horizons'', i.e., 
 they employ an {\em infinite} geodesic length to reach the  region at 
 $r=\infty$;
 obviously this kind of 
 geodesics could reach  the intersection 
 of the past and the future horizons at most, because of
  the time coordinate which remains costant. Exactly, the strange 
 behaviour of the above geodesics  
 arises from the fact  that the intersection of 
 the two horizons is not contained in the whole  manifold, in other words  
 {\em the future and the past horizons do not intersect at all}, and thus 
 these strange geodesics get lost into infinity.\\
 This is a well known feature of the extremal 
 Reissner-Nordstr\"{o}m space-time \cite{carter,wald libro,Hawking libro}.\\

 The geodesic equations of the metric (\ref{metrica}) are:
 \begin{eqnarray}
 \frac{d}{ds}\left( \frac{1}{r^{2}}\frac{dt}{ds}\right)&=&0 \label{auno}\\
\frac{d}{ds}\left( \frac{1}{r^{2}}\frac{dr}{ds}\right)&=&-\frac{1}{r^{3}}
\left[ \left(\frac{dr}{ds} \right)^{2}-\left(\frac{dt}{ds} \right)^{2} \right]
\label{adue}\\
\frac{d}{ds}\left( \frac{d\theta}{ds}\right)&=&\sin\theta\cos\theta
\left(\frac{d\phi}{ds} \right)^{2}\label{atre}\\
\frac{d}{ds}\left( \sin^{2}\theta \frac{d\phi}{ds}\right)&=&0 \label{aquattro}
 \end{eqnarray} 
 By an opportune choice of the affine parameter, posing $s$ to be equal 
 to the Riemannian
 length, we may add another equation to the previous  set of equations:
\beq
\frac{1}{r^{2}}\left[ \left(\frac{dr}{ds} \right)^{2}
-\left(\frac{dt}{ds} \right)^{2} \right]+\left( \frac{d\theta}{ds} \right)^{2}
+\sin^{2}\theta\left( \frac{d\phi}{ds} \right)^{2}=\pm1 \label{acinque} \:,
\eeq
where the right hand member is equal to $+1$ for space-like 
geodesics or $-1$ for
 time-like
geodesics.\\
Note that, because of the spherical symmetry, every geodesic defines by its
space-like components $r,\theta,\phi$ a spatial curve which
lies in
a space-like 3-plane which intersects the (spatial) origin. 
Indeed we can consider the geodesic variational equations
as the  {\em Euler-Lagrange} equations of motion   of a material point, 
by using the time $t$ as the evolution parameter.
Because of the spherical symmetry of the action, using  {\em Noether's}
 theorem,
one concludes that  there exists a space-like 3-vector, perpendicular 
to the spatial motion,
 which remains constant in  time. In fact from  Noether's theorem it
arises:
\beq
\ep_{ijk} \frac{\partial L}{\partial \dot{x}^{j}}
 x^{k}=\mbox{constant} \:,\label{lagrange}
\eeq
where we used the dot to indicate the time derivative.
Obviously we supposed also: 
\beq
L=\sqrt{\mid g_{00}+g_{pq}\dot{x}^{p}\dot{x}^{q}\mid } \:,\label{lagrangiana}
\eeq 
where: 
\beq
g_{00}= \frac{1}{r^{2}} \:\:\: \mbox{and} \:\:\: g_{pq}=
\frac{\delta_{pq}}{r^{2}}\:.\nonumber
\eeq
If we take the time derivative of (\ref{lagrange}) and  remember the 
Lagrangean form of the equation of motion we will obtain:
\beq
\ep_{ijk} \frac{\partial L}{\partial x^{j}}
 x^{k}+ \ep_{ijk} \frac{\partial L}{\partial \dot{x}^{j}}
 \dot{x}^{k}=0 \:. \nonumber
\eeq
It can easily be shown, using the antisymmetry of $\ep_{ijk}$
and the  explicit form of  the $\dot{x}^{j}-$derivative,
 that the second term of the left hand member vanishes.
It remains:
\beq
\ep_{ijk} \frac{\partial L}{\partial x^{j}}
 x^{k} =0 \:\: \: \mbox{i.e.} \:\:\: \frac{\partial L}{\partial \vec{x}} \wedge
\vec{x} =0\:. \nonumber
\eeq
As we argued this is the equation of a plane which intersects the origin.\\
 Once again, because of the spherical symmetry of
the problem,  we may consider only those geodesics which
lie in the plane $\theta=\pi/2$.\\
In this way the solutions of Eq.~(\ref{atre}) and Eq.~(\ref{aquattro})
are trivial:
\begin{eqnarray}
\theta&=&\frac{\pi}{2} \nonumber \:, \\
\phi&=&Bs+\phi_{0}\:,
\end{eqnarray} 
where $B$ and $\phi_{0}$ are real constants.
Consequently, we can write down  Eq.~(\ref{acinque}) as:
\beq
\frac{1}{r^{2}}\left[ \left(\frac{dr}{ds} \right)^{2}
-\left(\frac{dt}{ds} \right)^{2} \right]+
B^{2}=\pm1 \label{acinqueb} \:.
\eeq
After some trivial calculations and the
use of Eq.~(\ref{acinqueb}), we end up with 
the following four sets of solutions
  of 
 the above differential equations, 
 for all the possible geodesics of  the indicated plane 
which intersect  (or seem to do so) the horizons.\\

{\em Time constant geodesics which seem to reach the horizons.}
\begin{eqnarray}
r&=&r_{0}e^{\pm s} \:,\label{bzero}\\
t&=&t_{0} \:,\label{bzero1}\\
\phi&=& B s + \phi_{0} \:,\label{bzero2}
\end{eqnarray}
where $r_{0}>0$, $t_{0}$, $B$, $\phi_{0}$ are real numbers
 and $s$ is the geodesic 
length measured from the point  $(t_{0},r_{0},\theta=\pi/2,\phi_{0})$. 
Note that all these geodesics are space-like. 
We stress that these geodesics {\em seem} to reach the horizons 
as $s \rightarrow \pm
\infty$, but that is an obvious contradiction; actually, as discussed above, 
 this is a consequence of the fact  that $H^{+}\cap H^{-}=\emptyset$.\\

{\em First family of  geodesics which start from the horizons.}
\begin{eqnarray}
r&=&\frac{A}{ s}\:,\label{bunodegenere}\\
t&=&t_{0}\pm \frac{A}{s} \:,\label{bduedegenere}\\
\phi&=& s + \phi_{0}\label{btredegenere} \:,
\end{eqnarray}
where $A>0$, $t_{0}$, $\phi_{0}$ are real numbers. 
The origin of the parameter $s$ is
chosen in a way that  the starting
point of the geodesic is on the  horizon.
This horizon will be  $H^{+}$ if the sign in front of $A$
in Eq.~(\ref{bduedegenere})
is $+$, otherwise it will be  $H^{-}$ if the sign is $-$.\\
All these geodesics are {\em space-like}.\\

{\em Second family of  geodesics which start from  the horizons.}
\begin{eqnarray}
r&=&\frac{A}{\sin\left(\sqrt{K} s\right)}\:,\label{buno}\\
t&=&t_{0}\pm A\cot\left(\sqrt{K} s\right) \:,\label{bdue}\\
\phi&=& B s + \phi_{0} \label{baggiuntadue} \:,
\end{eqnarray}
where $K>0$, $A>0$, $B$, $t_{0}$, $\phi_{0}$ are real numbers; furthermore
$K$ and $B$ are related in only one of the following two possibilities:\\
 $K=1+B^{2}$ and the  geodesic is {\em time-like},\\
 or\\
 $B^{2}>1$, $K=B^{2}-1$ and the geodesic is  {\em space-like.}\\
The origin of the parameter $s$ is 
chosen in a way  that  the starting 
point of the geodesic is on the  horizon. 
This horizon will be  $H^{+}$ if the sign in front of $A$
in Eq.~(\ref{bdue})
is $+$, otherwise it will be  $H^{-}$ if the sign is $-$.\\

{\em Third family of geodesics which start from the horizon.}
\begin{eqnarray}
r&=&\frac{A}{\sinh\left(\sqrt{K} s\right) }\:,\label{btre}\\
t&=&t_{0}\pm A\coth\left(\sqrt{K} s\right)  \:,\label{bquattro}\\
\phi&=& B s + \phi_{0} \label{baggiuntatre} \:,
\end{eqnarray}
where $K>0$, $A>0$, $B$, $t_{0}$, $\phi_{0}$ are real numbers; furthermore
 $B^{2}<1$, $K=1-B^{2}$ and all geodesics are space-like.\\
 As in the previous case the 
 origin of the parameter $s$ is
chosen in a way  that  the starting
point of the geodesic is on the  horizon. This horizon
 will be $H^{+}$ if the sign in front of $A$
in Eq.~(\ref{bquattro})
is $+$, otherwise it will be $H^{-}$ if the sign is $-$.\\

Now we calculate the Wightman function for a massless scalar field in the 
metric
(\ref{metrica}).\\
This metric is conformal to the Minkowski metric, thus
 we may use the Dowker and Schofield's
method \cite{dowker1,dowker2} which connects the Wightman  functions 
(in general the Green functions) of a 
 scalar field in a static
manifold with the corresponding Wightman functions of the field in another,  
{\em conformally}
related static manifold.\\
Let us  suppose to have two {\em static} metrics which are 
conformally related:
\beq
 ds^{2}=g_{00}(\vec{x})(dx^{0})^{2}+g_{ij}(\vec{x})dx^{i}dx^{j} \label{cuno}
\eeq
and 
\beq
 ds^{'2}=g^{'}_{00}(\vec{x})(dx^{0})^{2}+g^{'}_{ij}(\vec{x})dx^{i}dx^{j}\:,
 \label{cdue}
\eeq
where
\beq 
g^{'}_{\mu\nu}=\lambda^{2}(\vec{x}) g_{\mu\nu}
\eeq
and let us  consider the Wightman functions which satisfy the 
respective Klein-Gordon equations:
\beq
\left(\Box +\xi R + m^{2}\right) W^{\pm}_{\be} (x,x^{'})=0  \label{KG1}
\eeq
and 
\beq
\left(\Box^{'} +\xi R^{'} +\left(\xi-\frac{1}{6}\right)\Box^{'}
\left(\lambda^{-2}\right)+  m^{2}\lambda^{-2}\right)
 W^{'\pm}_{\be} (x,x^{'})=0\:,  \label{KG2}
\eeq
where $R$ is the scalar curvature.\\
The Wightman functions above are related by the Dowker-Schofield scaling 
property:
\beq
 W^{'\pm}_{\be}(x,x^{'})= \lambda^{-1}(\vec{x}) W^{\pm}_{\be}(x,x^{'}) 
\lambda^{-1}(\vec{x^{'}}) \:, \label{scaling}
\eeq
In our case $ds^{'2}$ is the Minkowski metric (hence $R^{'}=0$),
  $ds^{2}$ is the metric 
of Eq.~(\ref{metrica2}) and thus $\lambda^{2}=\vec{x}^{2}$.\\
 The Wightman functions  which we are interested in satisfy the Klein-Gordon
 equation
in (\ref{KG1}) with $m=0$ and $R=0$, in fact the metric of 
 Eq.~(\ref{metrica2}) (but also the real extremal R-N metric)
 has vanishing scalar curvature. We stress that, due to above fact, 
the {\em minimal} coupling in Eq.(\ref{KG1}) 
coincides with the {\em conformal} coupling. 
 Thus we can choose the value of the parameter $\xi$
to be $\xi=1/6$ (conformal coupling) in  Eq.(\ref{KG1}) and hence 
in Eq.(\ref{KG2}) which reduces to the usual massless K-G equation 
 in the Minkowski space-time.\\
The thermal Wightman functions of a massless scalar field 
in Minkowki space can be  obtained 
 by using the procedure sketched in {\bf Section 1}. They read:
\beq
W^{\pm}_{M\be}=\frac{1}{4\pi^2}\frac{\pi \left\{\coth\frac{\pi}{\be}
(|\vec{x}-\vec{x}^{'}|+t-t^{'}\mp i\ep)+\coth
\frac{\pi}{\be}(|\vec{x}-\vec{x}^{'}|-t+t^{'}\pm i\ep)\right\} }{2\be
|\vec{x}-\vec{x}^{'}|} \label{wightman Minkowski}
\eeq 
Using Eq.(\ref{scaling}) we find the thermal 
Wightman functions of a massless scalar field 
(minimally as well as  conformally coupled)
 propagating outside of a large mass, extremal  
Reissner-Nordstr\"{o}m black-hole:
\beq
W^{\pm}_{\be}=   \frac{\mid \vec{x}\mid  \mid \vec{x^{'}}\mid}{4\pi^2}\frac{\pi \left\{\coth\frac{\pi}{\be}
(|\vec{x}-\vec{x}^{'}|+t-t^{'}\mp i\ep)+\coth
\frac{\pi}{\be}(|\vec{x}-\vec{x}^{'}|-t+t^{'}\pm i\ep)\right\} }{2\be
|\vec{x}-\vec{x}^{'}|} \label{wightman}
\eeq
If  we take the limit $\mid\vec{x}\mid\rightarrow +\infty$  we 
obtain the thermal Wightman functions calculated on the horizon 
(in the argument $x$). In order to  to calculate this limit we  must 
increase (to reach $H^{+}$) or decrease (to reach $H^{-}$)
the variable $t$ together the variable $\vec{x}$ and $\phi$ along the 
geodesics obtained above.
 Once again we  deal with
$H^{+}$ only because of the time symmetry of the problem.\\
 If we consider either 
the space-like
or the time-like geodesics of the above three families 
 we  will produce the same function:
\beq
W^{\pm}_{\be}\left(x_{H^{+}},x^{'} \right)=
\frac{r^{'}\pi \left\{ 1 + \coth\left[\frac{\pi}{\be}\left( t^{'}-t_{0}-r^{'}
\cos( \phi^{'}-\phi_{0})
\right) \right]   \right\} }{4\pi^{2} 2 \be}
\label{ddieci}
\eeq
Note that along the  curves satisfying:  
\beq
r^{'}\cos(\phi^{'}-\phi_{0}) =t^{'}-t_{0}\:, \nonumber
\eeq 
there is a divergence. Indeed it can be easily proved,
 starting from the metric of Eq.~(\ref{metrica}), that
 all the {\em light-like} geodesics
 which meet horizon $H^{+}$ satisfy this relation,
 thus this is the correct, expected divergence.\\

In order to check the HNS principle on $H^{+}$  for a space-like
 vector $z$ tangent to a  geodesic of the first family in the horizon,
we consider Eq.~(\ref{ddieci}) and
substitute  the variables $r^{'}$, $t^{'}$, $\phi^{'}$
for the   functions defined in the right hand side
of equations (\ref{bunodegenere}), (\ref{bduedegenere}), (\ref{btredegenere}),
using also the identifications: $t_{0}=t_{0}^{'} $ and $\phi_{0}=\phi_{0}^{'}$.
 Finally we redefine $s=\lambda z^{i}$.\\
We obtain, as $\lambda \rightarrow 0$:
\beq
\lambda^{2}W_{\be}\left(x_{H^{+}},x_{H^{+}}+\lambda z \right)\sim
\nonumber
\eeq
\begin{eqnarray}
&\sim&
\frac{A\pi \lambda^{2}}{4\pi^{2}2\be \lambda z^{i}}
\left\{ \coth\left[\frac{\pi}{\be}\left( A\frac{1}{ \lambda z^{i}}-
\frac{A \cos\left( \lambda z^{i}\right)}{\lambda z^{i}}\right)
\right] +1  \right\} \nonumber \\
&\sim&
\frac{A\pi \lambda^{2}}{4\pi^{2}2\be \lambda z^{i}}
\left\{ \sinh\left[\frac{\pi}{\be}\left( A
\frac{1}{\lambda z^{i}}-\frac{A\cos\left( \lambda z^{i}\right)}{\lambda 
z^{i}}\right)
\right] \right\}^{-1} \nonumber \\
&\sim&
\frac{A\pi \lambda^{2}}{4\pi^{2}2\be \lambda z^{i}}
\left\{ \frac{\pi}{\be}\left( A
\frac{1}{\lambda z^{i}}-\frac{A\cos
\left( \lambda z^{i}\right)}{\lambda z^{i}}\right)\right\}^{-1} \nonumber \\
&\sim& \frac{\lambda^{2}}{4\pi^{2}2 z^{i}}\frac{z^{i}}{\frac{1}{2}
(\lambda z^{i})^{2}}=\frac{1}{4\pi^{2}} \frac{1}{g_{\mu\nu}
\left( H^{+}\right) z^{\mu}z^{\nu}}\:. \nonumber
\end{eqnarray}
We conclude that in this case the HNS principle holds
 for any value of
$\be$. \\

In order to check the HNS principle on $H^{+}$ for a time-like 
 vector $z$ tangent to a geodesic of the  second
family on  the horizon,
we consider Eq.~(\ref{ddieci}) and 
substitute the  variables $r^{'}$, $t^{'}$, $\phi^{'}$ for the 
 functions defined in the right hand side
of equations (\ref{buno}), (\ref{bdue}), (\ref{baggiuntadue}), 
with the identifications: $t_{0}=t_{0}^{'} $ and $\phi_{0}=\phi_{0}^{'}$.
 Finally we redefine $s=\lambda z^{0}$.\\ 
We obtain, as $\lambda \rightarrow 0$:   
\beq
\lambda^{2}W_{\be}\left(x_{H^{+}},x_{H^{+}}+\lambda z \right)\sim
\nonumber
\eeq
\begin{eqnarray}
&\sim&
\frac{A\pi \lambda^{2}}{4\pi^{2}2\be \sin(\lambda \sqrt{K} z^{0})}
\left\{ \coth\left[\frac{\pi}{\be}\left( A \cot(\lambda \sqrt{K}z^{0})-
\frac{A \cos \left(B\lambda z^{0}\right)}{\sin(\lambda \sqrt{K}z^{0})}\right)  
\right] +1  \right\} \nonumber \\
&\sim&
\frac{A\pi \lambda^{2}}{4\pi^{2}2\be \sin(\lambda \sqrt{K}z^{0})}
\left\{ \sinh\left[\frac{\pi}{\be}\left( A 
\cot(\lambda\sqrt{K} z^{0})-\frac{A\cos\left( B\lambda z^{0}\right)}{\sin
(\lambda 
\sqrt{K}z^{0})}\right) 
\right] \right\}^{-1} \nonumber \\
&\sim&
\frac{A\pi \lambda^{2}}{4\pi^{2}2\be \sin(\lambda \sqrt{K}z^{0})}
\left\{ \frac{\pi}{\be}\left( A
\cot(\lambda \sqrt{K}z^{0})-\frac{A\cos \left( B\lambda z^{0}\right)}{\sin(\lambda 
\sqrt{K}z^{0})}\right)
 \right\}^{-1} \nonumber \\
&\sim&\frac{1}{4\pi^{2}2 \sqrt{K}z^{0}} 
\left\{ \frac{\cos(\lambda\sqrt{K} z^{0})-\cos(B \lambda z^{0})}{\lambda 
\sin(\lambda \sqrt{K}z^{0})} \right\}^{-1}\nonumber \\
&\sim&\frac{\lambda}{4\pi^{2}\sqrt{K}} \frac{\lambda \sqrt{K}}{\left(
\lambda B z^{0}\right)^{2}-\left(
\lambda \sqrt{K} z^{0}\right)^{2} }
\nonumber \\
&\sim&\frac{1}{4\pi^{2}}\frac{1}{z^{02}\left(B^{2}-K  \right)}
= \frac{1}{4\pi^{2}}\frac{1}{g_{\mu\nu}\left(H^{+}\right)
z^{\mu}z^{\nu}} \nonumber
\end{eqnarray}\\
In the last step we used $B^{2}-K=-1$.
We conclude that in this case the HNS principle holds 
 for any value of 
$\be$. \\
For space-like vectors the calculations are quite identical except for the fact
that one has to substitute $z^{0}$ for $z^{i}$ above and that it holds
$B^{2}-K=+1$ instead of $B^{2}-K=-1$.\\

In order to check the HNS principle on $H^{+}$  for a space-like
 vector $z$ tangent to a geodesic of the  third family on the horizon,
we consider Eq.~(\ref{ddieci}) and
we substitute the variables $r^{'}$, $t^{'}$, $\phi^{'}$
 for the functions defined in the right hand
of equations (\ref{btre}), (\ref{bquattro}), (\ref{baggiuntatre}),
with the identifications: $t_{0}=t_{0}^{'} $ and $\phi_{0}=\phi_{0}^{'}$.
 Finally we redefine $s=\lambda z^{i}$.\\
We obtain, as $\lambda \rightarrow 0$:
\beq
\lambda^{2}W_{\be}\left(x_{H^{+}},x_{H^{+}}+\lambda z \right)\sim
\nonumber
\eeq
\begin{eqnarray}
&\sim&
\frac{A\pi \lambda^{2}}{4\pi^{2}2\be \sinh(\lambda \sqrt{K} z^{i})}
\left\{ \coth\left[\frac{\pi}{\be}\left( A \coth(\lambda \sqrt{K}z^{i})-
\frac{A \cos\left( B\lambda z^{i}\right)}{\sinh(\lambda \sqrt{K}z^{i})}\right)
\right] +1  \right\} \nonumber \\
&\sim&
\frac{A\pi \lambda^{2}}{4\pi^{2}2\be \sinh(\lambda \sqrt{K}z^{i})}
\left\{ \sinh\left[\frac{\pi}{\be}\left( A
\coth(\lambda\sqrt{K} z^{i})-\frac{A\cos\left( B\lambda z^{i}\right)}{\sinh(
\lambda
\sqrt{K}z^{i})}\right)
\right] \right\}^{-1} \nonumber \\
&\sim&
\frac{A\pi \lambda^{2}}{4\pi^{2}2\be \sinh(\lambda \sqrt{K}z^{i})}
\left\{ \frac{\pi}{\be}\left( A
\coth(\lambda \sqrt{K}z^{i})-\frac{A\cos\left( B\lambda z^{i}\right)}{
\sinh(\lambda
\sqrt{K}z^{i})}\right)
 \right\}^{-1} \nonumber \\
&\sim&\frac{1}{4\pi^{2}2 \sqrt{K}z^{i}}
\left\{ \frac{\cosh(\lambda\sqrt{K} z^{i})-\cos(B \lambda z^{i})}{\lambda
\sinh(\lambda \sqrt{K}z^{i})} \right\}^{-1}\nonumber \\
&\sim&\frac{\lambda}{4\pi^{2}\sqrt{K}} \frac{\lambda \sqrt{K}}{\left(
\lambda B z^{i}\right)^{2}+\left(
\lambda \sqrt{K} z^{i}\right)^{2} }
\nonumber \\
&\sim&\frac{1}{4\pi^{2}}\frac{1}{z^{i2}\left(B^{2}+K  \right)}
=\frac{1}{4\pi^{2}}\frac{1}{g_{\mu\nu}\left(H^{+}\right)
z^{\mu}z^{\nu}} \nonumber
\end{eqnarray}\\
In the last step we used $B^{2}+K=+1$.
We conclude that in this case the HNS principle holds
 for any value of
$\be$. \\
We finally stress that the HNS principle accepts the limit value of the
 temperature $T=0$, too.
 Indeed, starting from Wightman functions in
Minkowski space-time
we find the {\em zero temperature} Wightman functions
in our coordinate by using
 Eq.~(\ref{scaling}). They read as:
\beq
W^{\pm}(x,x^{'})=\frac{\mid \vec{x}\mid  \mid \vec{x^{'}}\mid}{4\pi^2}
\frac{1}{\sigma^{2}(x_{\mp\ep},x')}\nonumber\:.
\eeq
The same result arises doing the limit as 
$\beta \rightarrow +\infty$ in Eq.(\ref{wightman}).\\ 
By putting an argument on the horizon we obtain:
\beq
W^{\pm}\left(x_{H^{+}},x^{'} \right)=
\frac{1}{8\pi^{2}} \left( \frac{t^{'}-t_{0}}{r^{'}}
-\cos( \phi^{'}-\phi_{0})\right)^{-1}\:.
\label{ultimaHNS}
\eeq
Once again we may observe that one obtains the same result doing the limit as 
$\beta \rightarrow +\infty$ in Eq.(\ref{ddieci}) directly.\\
Now it is really very easy to check, following the usual procedure, 
that the above
Wightman functions satisfy the HNS principle 
on the  horizons for all  time-like
or space-like  geodesics which reach the horizons and they 
diverge for light-like
ones.\\

Finally, let us check the {\em Hessling principle} in the present case.\\
We consider the space-like geodesics of the  first family in Eq.s 
(\ref{bunodegenere}), (\ref{bduedegenere}) and  (\ref{btredegenere}).
It arises by using that family of geodesics and  Eq.(\ref{ddieci}):
\beq
\frac{d\:\:}{d\lambda} \:
 \lambda^{2} W_{\be}(x_{H^{+}},x_{H^{+}}
 +\lambda z)  =  \frac{\Gamma}{\beta}\: 
\frac{d\:\:}{d\lambda} \: 
 \lambda 
\left\{ 1+ \coth \left[\frac{\pi A}{\beta \lambda z} (1- \cos \lambda z)
\right] \right\} \:, \nn
\eeq
where the factor $\Gamma:= A/8\pi z$ is {\em finite}
 and it does not depend on $\lambda$ and $\beta$.\\
Some trivial calculations lead to:
\beq
\frac{d\:\:}{d\lambda} \:
 \lambda^{2} W_{\be}(x_{H^{+}},x_{H^{+}}
 +\lambda z)  =  \frac{\Gamma}{\beta}
\left\{ 1+ \frac{X + \sinh X\: \cosh X - 
\frac{\pi A}{\beta z} \sin \lambda z}{\sinh^{2}X}  \right\}\:, \nn
\eeq
where we also posed:
\beq
X := \frac{\pi A}{\beta z} \frac{1- \cos \lambda z}{\lambda}\:. \nn
\eeq
Expanding around $\lambda=0$, i.e., $X=0$ it arises:
\beq
\frac{d\:\:}{d\lambda} \:
 \lambda^{2} W_{\be}(x_{H^{+}},x_{H^{+}}
 +\lambda z)  =  \frac{\Gamma}{\beta}
\left\{ 1+ \frac{ O(\lambda^{3})}{\lambda^{2}+O(\lambda^{4})}\right\}\:, \nn
\eeq
Hence we get, doing the limit as $\lambda \rightarrow 0^{+}$:
\beq
\frac{d\:\:}{d\lambda} \:
 \lambda^{2} W_{\be}(x_{H^{+}},x_{H^{+}}
 +\lambda z) \rightarrow \frac{\Gamma}{\beta}\:.\nn
\eeq 
{\em This fact  is sufficient to prove that the Hessling principle excludes 
every
finite value of $\beta$}.\\
The limit case $T=1/\beta= 0$ survives only. Using the remaining two families
of geodesics, calculations result to be very similar and the same limit value
of the temperature survives. Furthermore, 
 it can be  simply  proved that the Wightman functions
of the R-N vacuum, i.e., the limit case $T=1/\be=0$, satisfies the Hessling
principle by considering directly
Eq.(\ref{ultimaHNS}). In fact it follows from Eq.(\ref{ultimaHNS})
 using geodesics
of the first family:
\beq
\frac{d\:\:}{d\lambda} \:
\lambda^{2} W(x_{H^{+}},x_{H^{+}}
 +\lambda z) =\nn 
\eeq
\beq
=\frac{1}{8\pi^{2}}
\frac{2\lambda -2\lambda \cos(\lambda z^{i})\:- z^{i}\lambda^{2}
\sin(\lambda z^{i}) }{(1- (\lambda z^{i}))^{2}}
=\frac{1}{8\pi^{2}}\frac{O(\lambda^{5})}{O(\lambda^{4})} \rightarrow 0
\:\:\:\: \mbox{as}\:\: \lambda \rightarrow 0^{+}\:.\nn
\eeq
By using the geodesic of the second family we obtain similarly:
\beq
\frac{d\:\:}{d\lambda} \:
\lambda^{2} W_{\be}(x_{H^{+}},x_{H^{+}}
 +\lambda z) = \nn
\eeq
\beq
 =\frac{1}{8\pi^{2}}\frac{2\lambda}{\cos(\lambda z \sqrt{K})
-\cos(\lambda z B)}+
\frac{\lambda^{2} (\sqrt{K}z \sin (\lambda z \sqrt{K})  
 - B z \sin (\lambda z B))}{(\cos(\lambda z \sqrt{K})
-\cos(\lambda z B))^{2}}=\nn
\eeq
\beq 
=\frac{\lambda O(\lambda^{4})}{O(\lambda^{4})} \rightarrow 0 
\:\:\:\: \mbox{as}\:\: \lambda \rightarrow 0^{+} \:.
\eeq
Finally, using the third family of geodesics:
\beq
\frac{d\:\:}{d\lambda} \:
\lambda^{2} W_{\be}(x_{H^{+}},x_{H^{+}}
 +\lambda z) = \nn
\eeq
\beq
 =\frac{1}{8\pi^{2}}\frac{2\lambda}{\cosh(\lambda z \sqrt{K})
-\cos(\lambda z B)}
-\frac{\lambda^{2} (\sqrt{K}z \sinh (\lambda z \sqrt{K})
 + B z \sin (\lambda z B))}{(\cosh(\lambda z \sqrt{K})
-\cos(\lambda z B))^{2}}= \nn
\eeq
\beq
=\frac{\lambda O(\lambda^{4})}{O(\lambda^{4})} \rightarrow 0
\:\:\:\: \mbox{as}\:\: \lambda \rightarrow 0^{+}  \:.
\eeq


\section{Discussion}

The most  important  conclusion which follows from the above
calculations  is that  the {\em weak} HNS principle,
 in the case of an extreme Reissner-Nordstr\"{o}m
black hole, holds for every value of $\be$, 
 i.e., once again it agrees with 
the other method based on the 
 elimination of the singularities of the  Euclidean manifold, but
 the {\em weak} Hessling principle selects only the  null
 temperature, i.e., the R-N vacuum as a physically 
sensible state.\\
We observe that our ``along geodesics'' calculations 
 eliminate quantum states which do not have the correct scaling limit,
on the other hand 
 one can not correctly think that they determine  only the states which
 have the correct
scaling limit in the sense precised in \cite{hessling}.
 In fact we used a {\em weaker} prescription as previously 
stressed.\\  
Another important point is that we dealt with
  the limit of a large mass black
 hole and with a massless field, but we think that our conclusions
 should hold without to assume these strong conditions, too.\\
 Indeed, in order to check the behaviour of Wightman functions on the
 horizon we recognize that is sufficient to know the form of the
 Wightman functions {\em near the horizon} only.
 In this region,
 regardless  of the value of  black hole's  mass, the metric can be
 written  in the form used above:
 \beq
 ds^{2}\sim \frac{1}{r^{2}}\left[ -dt^{2}+dr^{2}+r^{2}d\Om_{2}\right]\:,  
 \nonumber
 \eeq
  thus  we expect that the Wightman functions for  arguments
 near
 the horizon 
 should be of the form (\ref{wightman}) and thus it should
  be possible to restore all our results.
 For example, we stress that Haag, Narnhofer
 and Stein in \cite{haag} used just the limit form of the metric near the
 horizon in order to obtain the Hawking temperature. 
 However, one could object that the normalization of the  modes used
  to construct the Wightman functions  depends on the integration over
 the
 whole spatial manifold and not only on the region near the horizon.
 Really, it is possible to
 overcome this problem at least formally dealing with
 our static metric.  In fact, in this case one recovers by the KMS
 condition
 \cite{haag} ($x \equiv (\tau, \vec{x})$):
 \beq
 < \phi(x_{1}) \phi(x_{2}) >_{\beta}\: = \frac{i}{2\pi}
 \int_{-\infty}^{+\infty} G(\tau_{1}+\tau,
 \vec{x}_{1} \mid \tau_{2}, \vec{x}_{2})
 \frac{e^{\be \om}}{e^{\be \om}-1} e^{i\om \tau}\: d\tau d\om
 \label{unohaag} \:,
 \eeq
 where the distribution $G$ is the {\em commutator} of the fields
 and thus it is uniquely determined \cite{haag} by the fact that
 it is a solution  of the Klein-Gordon equation in both arguments,
 vanishes for equal times $\tau_{1} = \tau_{2}$ and is normalized by
 the ``local'' condition:
\beq
 g^{\tau\tau}\sqrt{-g}\frac{\partial}{\partial \tau_{1}}
  G(x_{1}, x_{2}) \mid_{\tau_{1}=\tau_{2}}
 = \delta^{3}(\vec{x}_{1}, \vec{x}_{2}) \label{duehaag}
 \eeq
 The above $3$-delta function is usually understood as:
 \begin{eqnarray}
  \delta^{3}(\vec{x}_{1}, \vec{x}_{2})&=&0 \:\:\:\: \mbox{for } \:\:\:\:
  \vec{x}_{1} \neq \vec{x}_{2} \nn\\
  \int \delta^{3}(\vec{x}_{1}, \vec{x}_{2})\:\: d\vec{x}_{2} &=& 1 \nn
 \end{eqnarray}
 By the previous, spatially ``local''
  formulas  we expect that the function $G$ calculated by using
 the ``true'' static metric becomes the function $G$ calculated by using
 the  approximated static  form of the metric  
 inside of a certain
 static region
 $\delta\Sigma \times \R$ (where  $ \tau \in \R$) as $\delta\Sigma$ shrinks
 around a $3$-point. Really, considering 
 $(\delta\Sigma,\tau_{0})$ as a 
{\em Cauchy surface},
 the above result should come out  inside of the ``diamond-shaped'' $4-$region
 causally determined by $(\delta\Sigma,\tau_{0})$ at least. But, studying
 the form of the light cones near event horizons of the
 form $|\vec{x}|=r_{0}$, $\tau \in \R$
 (including the limit case case $r_{0}\rightarrow \pm \infty$),
  it is simple to prove
 that this $4-$region will tend to 
 contain  the whole $\tau-$axis if $\delta\Sigma$
 approachs to the event horizons.\\ 
 This is  the case of an extremal R-N black hole where 
 the Bertotti-Robinson
 metric approximates the R-N metric near the horizon $r\rightarrow +\infty $,
 $t\in \R$.\\
 In the same way, using Eq.~(\ref{unohaag}), we could expect such a
 property for thermal Wightman functions, too, the case
 of zero temperature, which is regarded as the limit
 $\be \rightarrow +\infty$, included.
 Furthermore, if  the field's mass $m$ in Eq.~(\ref{KG1}) were not
 zero (and we chose again $\xi=1/6$ following the previous motivations), 
 the Wightman functions which we are
 interested in would by be connected to the
 Wightman functions satisfying the ordinary Klein-Gordon equation in the
 Minkowski space-time except for a 
  position dependent mass term
 (see Eq.~(\ref{KG2})):  
\beq
M^{2}(r)=\frac{m^{2}}{r^{2}} \:. \nonumber
\eeq 
We stress that this mass term is
  vanishing as $r\rightarrow +\infty$. 
Our method  works for large value of $r$ so we expect that our
conclusions do not change in the case of a {\em massive} scalar field.\\
Finally, we stress that recently P.H. Anderson, W.A. Hiscock and D.J 
Loranz \cite{AHL}, by using of the metric in Eq.(\ref{metrica}) and the 
Brown-Cassidy-Bunch formula (see \cite{AHL,rindler} and references therein) 
argued
(and numerically checked by using the {\em complete}
 R-N metric) that the 
Reissner-Nordstr\"{o}m vacuum state is the {\em only} thermal
state with a non-singular renormalized stress-tensor on the horizon of an
extremal R-N black hole. In fact they obtained the formula holding
near the horizon:
\beq
< T_{\mu}^{\nu} >_{ \be renorm.} \sim \frac{1}{2880 \pi^{2} } 
\delta^{\mu}_{\nu} + r^{4} \frac{\pi^{2}}{30 \be^{4}} \:\mbox{diag}\left(
-1,\frac{1}{3},\frac{1}{3},\frac{1}{3} \right) 
\eeq
Note that, if $T=1/\be$  does not vanish, there will be 
 a strong divergence as $r \rightarrow +\infty $, i.e., on 
the horizon.
This divergence is not due to the coordinate frame used because it remains
also for scalar quantities as $T_{\mu}^{\nu}T_{\nu}^{\mu}$.
Supposing the stress-tensor generates the gravity by means of Einstein's
equations (or by some similar generalisation), the above divergence generates
a singularity in the metric structure of the  manifold.
 In the framework of the {\em Semiclassical Quantum
Gravity} (see \cite{rindler} for example)   the R-N vacuum state
results to be the only  possible state in equilibrium with an extremal
 R-N black hole.\\  
 We observe that  the ``improved''  HNS prescription, i.e. the 
 Hessling principle agrees completely with the result of 
Anderson Hiscock and Loranz  in our weaker formulation at least,
 in particular it selects a state carrying 
a renormalized stress-tensor  finite on the horizon.
This fact comes out also both in the Rindler space where the HNS and 
Hessling's 
prescriptions selects the Minkowski vacuum which has a regular stress tensor
on the horizon or in the Schwarzschild space where the HNS principle
selects the Hartle-Hawking state with the same property on the horizon
\cite{rindler}.


\ack{I would like to thank G.Cognola, M.Toller, L.Vanzo 
 and S.Zerbini of the Dipartimento di Fisica dell'Universit\`{a} di 
 Trento for all the useful discussions about the topics of this  paper.\\
 I am  grateful to Bernard Kay and a referee of
 {\em Classical and Quantum Gravity}
  who pointed out the Hessling paper to me.\\  
 Finally, I would like to thank S. Steidl of the  Physics Department of
 the  University of Innsbruck for the constant and patient technical 
 assistance.}


\newpage

\begin{thebibliography}{6}

\bibitem{H-H} Hawking~S~W, Horowitz~G~T, Ross~S~F 1995
Entropy, Area, and Black Hole Pairs
{\em Phys.Rev.} {\bf D 51} 4302,
{\em  gr-qc}/9409013

\bibitem{prolungamento} Narlikar~J~V and  Padmanabhan~T  1986 {\em Gravity,
Gauge Theories and Quantum Cosmology} (Dordrecht: D.Reidel Publishing Company)

\bibitem{kms} Kubo R 1957
{\em J. Math. Soc. Jpn.} {\bf 12} 570,\\
\\
Martin P C Schiwinger J 1959
{\em Phys. Rev.} {\bf 115} 1342,\\
\\
Haag R, Hugenholtz N M and Winnink M 1967
  {\em Commun.~Math.~Phys} {\bf 5} 215

\bibitem{libro haag} Haag~R 1992 {\em Local Quantum Physics}
(Berlin: Springer-Verlag)

\bibitem{gorini} Bertola M, Gorini V, Zeni M 1995
Quantized Temperature Spectra in Curved Spacetime 
{\em preprint Universita' di Milano} DYSCO 056,IFUM 513/FT
{\em hep-th}/9508004

\bibitem{haag} Haag~R, Narnhofer~H, Stein~U 1984
{\em Commun. Math. Phys.} {\bf 94} 219

\bibitem{FH} Fredenhagen K Haag R 1987
{\em Commun. Math. Phys.} {\bf 108} 91


\bibitem{hessling} Hessling H 1994 
{\em Nuc. Phys.}  {\bf B 415} 243
 
\bibitem{wald} Kay~B~S and Wald~R~M 1991 {\em Phys. Rep.} {\bf 207} 49

\bibitem{rindler} Birrel~N~D and Davies~P~C~W 1982 {\em Quantum Field Theory in
  Curved Space}.   (Cambridge: Cambridge University Press)

\bibitem{fulling} Fulling~S~A 1989 {\em Aspects of Quantum Field 
Theory in Curved
Space-Time.} (Cambridge: Cambridge University Press)

\bibitem{fr}
S.A.~Fulling and S.N.M.~Ruijsenaars.
 Phys.~Rep. {\bf 152}, 135 (1987).

\bibitem{dowker1}  Kennedy~G, Critchley~R~C, Dowker~J~S 1980 {\em Ann. Phys.}
{\bf 125} 346

\bibitem{dowker3}Dowker J S 1977
 {\em J.~Phys.} {\bf A10} 115 

\bibitem{dowker4}Dowker J S 1978
 {\em Phys.~Rev.} {\bf 18} 1856 

\bibitem{conic0} Dowker~J~S 1987 
{\em Phys. Rev.} D {\bf 36} 3095 

\bibitem{conic-1} Hawking S W 1978
{\em Phys. Rev.}{\bf D16} 1747

\bibitem{conic1} Dowker~J~S 1987
{\em Phys. Rev.} D {\bf 36} 3742 

\bibitem{conic2} Cognola~G, Kirsten~K, Vanzo~L 1994
{\em Phys. Rev.} D {\bf 49} 1029 

\bibitem{conic3} Solodukhin~G 1995 The Conical Singularity and
Quantum Corrections to BH Entropy,
{\em Phys.Rev.}{\bf D 51} 609, 
{\em hep-th}/9407001 

\bibitem{giapponese} Takagi~S 1986 {\em Progr. Theor. Phys. Suppl.} {\bf 88} 
 
\bibitem{cvz} Cognola~G, Vanzo~L, Zerbini~S 1995 One-loop Quantum
Corrections to the Entropy for an Extremal Reissner-Nordstr\"{o}m Black-Hole
 {\em Phys.Rev.}{\bf D 52} 4548, {\em  hep-th}/9504064

\bibitem{bertotti robinson} Bertotti~B 1959 {\em Phys. Rev.} {\bf 116} 1331,\\
\\
 Robinson~I 1959 {\em Bull. Acad. Pol. Sci.} {\bf 7} 351
                              
\bibitem{carter} Carter~B 1996 {\em Phys. Rev. Lett.} {\bf 21} 423

\bibitem{dowker2} Dowker~J~S and  Schofield~J~P 1988 
{\em  Phys. Rev.} D {\bf 38} 3327

\bibitem{wald libro} Wald~R~N 1984 {\em General Relativity} (Chicago:
 The University of Chicago Press)


\bibitem{Hawking libro} Hawking~S~W and Ellis~G~F~R 1973 {\em The Large Scale
Structure
of the Space-Time.} (Cambridge: Cambridge University Press)

\bibitem{AHL} Anderson~P~R, Hiscock~W~A, Loranz~D~J 1995 
{\em Phys. Rev. Lett.} {\bf 74} 4365, {\em gr-qc}/9504019




\end{thebibliography}
\end{document}